# Where does pancreatic lesion segmentation fail on public CT? An attenuation-stratified analysis of a duct-guided pipeline on PanTS

Kang-Won Byun (Independent Researcher), bgw2001@gmail.com

## Abstract

**Background.** Lesion segmentation is a building block of most CT pipelines for pancreatic cancer detection, yet lesion overlap scores stay low. Instead of proposing a new architecture, we report where an end-to-end pipeline fails on a large public dataset.

**Methods.** On PanTS we built a three-stage pipeline: pancreas segmentation, maps of the main pancreatic duct (MPD) and common bile duct (CBD), and a lesion model that receives the CT with or without the predicted duct maps. Both lesion models were trained identically on 4007 scans (417 with a lesion), selected on 150 validation scans, and evaluated once on 737 held-out test scans (127 with a lesion) under a comparison rule fixed before the test run. Out-of-fold predictions on 168 training scans with a lesion and 1433 without were used to locate failures: oracle removal of false components, lesion coverage, lesion attenuation relative to normal pancreas, candidate filtering, and MPD and CBD diameters as standalone signs of lesion presence.

**Results.** On the test set, lesion Dice was 0.213 (95% CI 0.173 to 0.259) with duct maps and 0.209 (95% CI 0.169 to 0.252) with CT only (paired difference +0.004 (95% CI -0.011 to +0.019)). The duct maps raised the share of lesions touched by a prediction (+0.079 (95% CI +0.024 to +0.134)) but also enlarged predictions in lesion-free scans. Lesion-like regions were predicted in 100.0% of lesion-free test scans. Out of fold, keeping only the predicted components that touched the lesion raised Dice from 0.285 to 0.309. The touching component covered a median 56.4% of the lesion, and a median 29.0% of its volume was lesion. Clearly hypoattenuating lesions, which intensity alone separates from normal pancreas, had the lowest coverage (median 16.6%). A 3D CNN on candidate crops did not outperform the component volume as a candidate filter. From the predicted maps alone, the MPD diameter separated lesion from lesion-free test scans with AUC 0.794 (95% CI 0.747 to 0.839); the CBD diameter was weak (0.603 (95% CI 0.549 to 0.658)) and adding it did not help (difference -0.000 (95% CI -0.009 to +0.010)). An MPD diameter of at least 5 mm had test sensitivity 44.1% and specificity 95.9%.

**Conclusions.** In this setting low lesion Dice came mainly from the extent of correctly located predictions and from predictions in lesion-free scans, not from extra components in lesion scans. Out of fold, about a third of lesions had the same attenuation as normal pancreas, which limits what any intensity-based boundary can achieve. Code for training, inference and evaluation is publicly available.

## 1. Introduction

Pancreatic ductal adenocarcinoma (PDAC) is often detected late, and contrast-enhanced CT is the main imaging test. Many automated detection systems first segment the pancreas and candidate lesions and then classify each candidate; ePAI, for example, segments the pancreas and surrounding anatomy, localizes all potential lesions, and classifies each lesion as PDAC, non-PDAC or normal [1]. Secondary signs such as MPD dilation are known to matter: iso-attenuating tumors "are identified only by the presence of secondary imaging findings (such as the dilation of the pancreatic duct)" [2], and duct and CBD masks have been used as additional network inputs [3].

Lesion segmentation quality on large public data is less well characterised, and low overlap scores are often attributed to small or iso-attenuating lesions without a breakdown of where the error comes from. Large public datasets with voxel labels for the pancreas, MPD, CBD and lesions now make such a breakdown possible.

This report has four parts: 1. an end-to-end pipeline on PanTS (pancreas → MPD/CBD maps → lesion), with a lesion model given the predicted duct maps and an identically trained CT-only control, compared once on a held-out test set under a pre-specified rule; 2. a decomposition of the lesion Dice loss into false components, missed lesions, and the extent of correctly located predictions; 3. an analysis stratified by lesion attenuation relative to normal pancreas and by lesion size; 4. negative results for candidate filtering (a 3D CNN on crops versus the component volume) and an analysis of MPD and CBD diameters as standalone signs of lesion presence.

## 2. Related work

ePAI uses a three-stage cascade of anatomy segmentation, lesion localization and lesion classification [1]. Alves et al. segment the pancreas and surrounding anatomy, including the duct, with nnU-Net and report a case in which an iso-attenuating tumor was located "following the termination of the dilated duct" [2]. Viviers et al. "stack the CT scan, pancreas and two ducts channel-wise" as input for tumor segmentation and detection [3]. Detection of direct and indirect imaging findings [4] and of MPD dilatation [5] has also been reported; for these two we read only the abstracts. Our pipeline therefore does not claim a new architecture. Its use of predicted duct maps follows [3].

## 3. Data

**PanTS.** We used the PanTS CT collection with its voxel labels for the pancreas, MPD, CBD and one "pancreatic lesion" class [6]. PanTS covers pancreatic ductal adenocarcinoma, pancreatic neuroendocrine tumors, pancreatic cystic neoplasms and cystic non-neoplastic lesions [6], and the lesion label carries no lesion type, so "lesion" below means any labelled pancreatic lesion. Scans that also appear in the PANORAMA collection (pixel-verified duplicates) were removed before splitting.

**Split.** Train 4007 scans (417 with a lesion), validation 150 (16), test 737 (127). The test set is the PanTS ImageTe partition; training and validation scans were drawn from ImageTr. PanTS provides no patient identifier, so the split is by scan; repeated scans of one patient across splits cannot be excluded. The test set had been used once before this study to compare duct segmentation models; it had not been used for any lesion-model decision.

## 4. Methods

### 4.1 Pipeline

*Stage 1, pancreas.* A 3D SegResNet [7] (MONAI implementation [8]) trained on the training scans segments the pancreas on the full CT. A region of interest (ROI) is the predicted pancreas plus a 30 mm margin, resampled to $0.8 \times 0.8 \times 1.0$ mm.

*Stage 2, duct maps.* A second SegResNet, trained on training scans with a dilated MPD, predicts the MPD and the CBD inside the ROI; maps are thresholded at 0.5.

*Stage 3, lesion.* Two lesion models with the same SegResNet configuration (16 initial filters, encoder blocks 1-2-2-4) predict two sigmoid outputs, the pancreas (including the lesion) and the lesion:

- *CT only (control):* input is the CT, intensities clipped to −175…250 HU and scaled to [0, 1];
- *with duct maps:* the CT plus the two binary maps predicted by stage 2.

Training was identical for both: $160 \times 160 \times 128$ patches sampled with label-class ratios (background 1, pancreas 1, MPD 1, CBD 0.5, lesion 1.5), ROI-margin jitter of 15 mm, Dice + cross-entropy loss with batch Dice, AdamW (learning rate $2 \times 10^{-4}$), batch 2, 250 iterations per epoch. A cosine schedule planned for 300 epochs was stopped after 100 epochs for both models, and the checkpoint with the best validation lesion Dice was kept (epoch 95 for both). Validation lesion Dice was 0.276 with duct maps and 0.271 for the control (16 lesion scans). Inference uses a sliding window; a voxel is lesion if the sigmoid exceeds 0.5, and predicted components are 26-connected.

Figure 1 shows the three stages on one test scan.

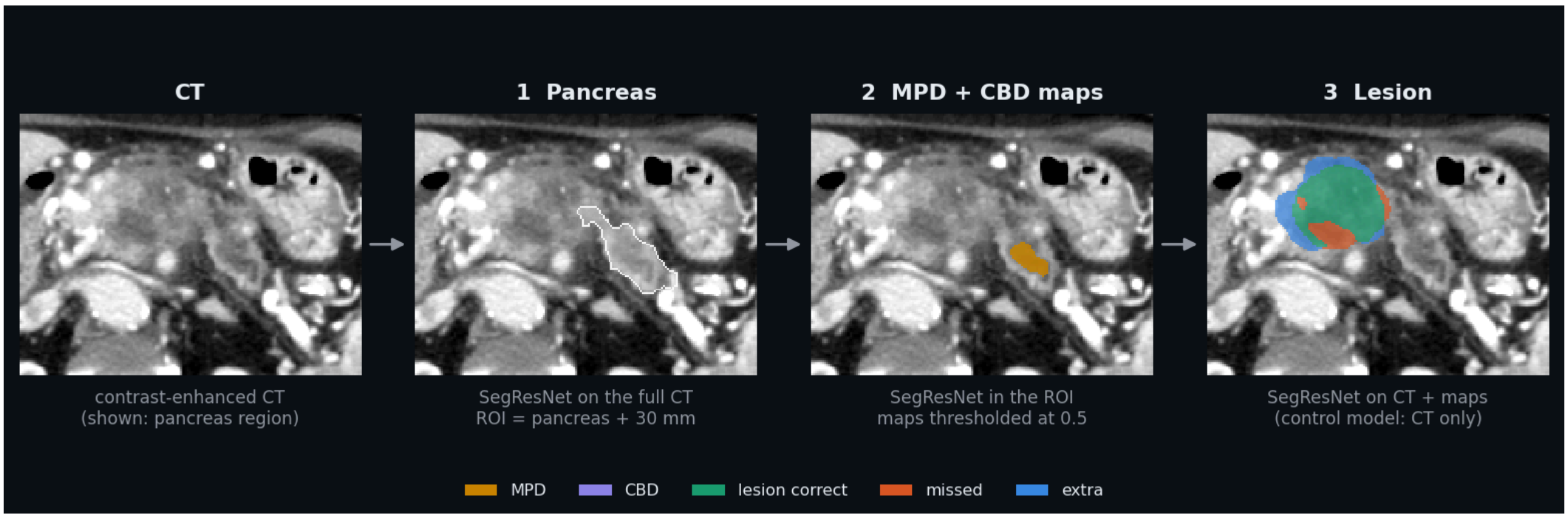

**Figure 1.** The pipeline on one test scan (case A of Figure 3): the CT (pancreas region shown), the stage-1 pancreas, the stage-2 MPD and CBD maps, and the stage-3 lesion prediction of the model with duct maps compared with the reference (green: correct, orange: missed, blue: extra prediction).

### 4.2 Out-of-fold predictions for the failure analysis

The training scans were split into five folds stratified by lesion presence. Two fold models (duct-map model, cosine schedule completed within 100 epochs, otherwise as above) were trained, and each scan of folds 0 and 1 was predicted by the model that had not seen it, through the full pipeline: 168 lesion scans and 1433 lesion-free scans. The remaining three folds were not trained, for time reasons. Stage 1 and stage 2 models had been trained on training scans, so their outputs on these scans are in-sample.

### 4.3 Failure analyses (out of fold)

- *Oracle component removal:* lesion Dice when every predicted component that does not touch the reference lesion is deleted, i.e. the ceiling of any component-level filter.
- *Coverage and lesion share:* for the components touching the lesion, the share of the lesion they cover and the share of their volume that is lesion.
- *Excess-removal ceiling:* lesion Dice if the touching components kept only their overlap with the lesion.
- *Probability threshold:* lesion Dice at thresholds 0.5 to 0.99 (fold 0).
- *Attenuation strata:* for each lesion, the probability that a random lesion voxel is darker than a random voxel of normal pancreas (voxel AUC). Normal pancreas = predicted pancreas eroded by one voxel, minus the lesion dilated by two voxels, minus the reference MPD/CBD dilated by one voxel. Lesions were labelled clearly brighter (AUC ≤ 0.4), similar (0.4–0.6), darker (0.6–0.8) or clearly darker (> 0.8). The same AUC was computed inside the touching components, lesion part versus the rest (both parts ≥ 20 voxels).
- *Size strata:* lesion-volume tertiles of the out-of-fold lesion scans (2.15 and 10.06 ml), applied unchanged to the test set.
- *Candidate filtering:* each predicted component was scored by (i) its volume, (ii) gradient boosting on 30 component measurements (size, probability, shape, location, intensity relative to the pancreas, distance to the duct maps), and (iii) a 3D ResNet-10 (MONAI implementation [8]) trained from scratch on a CT crop of the component bounding box plus 10 mm, resampled to $64^3$ voxels. Classifiers were cross-fitted between folds 0 and 1. Operating points were set to find 80% of lesion scans.
- *Duct and CBD diameters (added after the main test run):* the maximum local diameter (twice the distance transform) of the predicted MPD map and of the predicted CBD map, each as a scan-level score for lesion presence (lesion yes or no; lesion type is not available). The two diameters were combined by a logistic regression (standardised inputs, default settings, no tuning) fitted once on all out-of-fold training scans and applied once to the test set; whether adding the CBD helped was judged by the paired test AUC difference with the same bootstrap rule. As a check for a shortcut, the diameters were recomputed after removing the predicted map voxels inside the reference lesion (dilated by one voxel). Fixed cutoffs and lesion-size strata were reported without fitting.

### 4.4 Test protocol and statistics

The test evaluation was specified and committed before the run: the two lesion models and their checkpoints, the evaluation script and the 0.5 threshold were those used on validation, and the main comparison was the paired difference (duct maps − control) in lesion Dice, lesion found rate and scan-level AUC of the maximum lesion probability. A difference was called an improvement only if its 95% bootstrap confidence interval (2,000 case resamples, seed 0) excluded 0. Nothing was selected on the test set. Attenuation and size strata on the test set are descriptive. Scans whose stage-1 pancreas prediction was empty are reported separately.

## 5. Results

### 5.1 Test set: duct maps versus CT only

Table 1 and Figure 2 list the test results. Lesion Dice was 0.213 (95% CI 0.173 to 0.259) with duct maps and 0.209 (95% CI 0.169 to 0.252) with CT only; the paired difference was +0.004 (95% CI -0.011 to +0.019), the found-rate difference +0.079 (95% CI +0.024 to +0.134) and the scan-level AUC difference -0.022 (95% CI -0.064 to +0.022). By the pre-specified rule, the duct maps did not change lesion Dice or scan-level AUC; they raised the found rate. The higher found rate came with more prediction in lesion-free scans (median predicted volume 15.716 ml with duct maps versus 12.165 ml with CT only). Both models predicted a lesion somewhere in nearly all lesion-free scans (100.0% and 99.8%). The maximum lesion probability was at least 0.999 in 95.3% of lesion scans and 98.7% of lesion-free scans with CT only (97.6% and 99.8% with duct maps), so it carried almost no scan-level information (AUC 0.468 and 0.446); the predicted volume did slightly better (AUC 0.587 and 0.545). One lesion-free scan failed because the stage-1 pancreas prediction was empty and one had no reference pancreas label and was excluded at preprocessing, leaving 127 lesion scans and 608 lesion-free scans.

**Table 1.** Test set (737 scans; 127 lesion and 608 lesion-free scans evaluated).

| | CT only | With duct maps |
|---|---|---|
| Lesion Dice, mean (95% CI) | 0.209 (95% CI 0.169 to 0.252) | 0.213 (95% CI 0.173 to 0.259) |
| Lesion Dice, median | 0.079 | 0.084 |
| Lesion found (prediction overlaps lesion) | 66.9% | 74.8% |
| Lesion covered by prediction, median | 24.5% | 32.4% |
| Lesion-free scans with any prediction | 99.8% | 100.0% |
| Predicted volume per lesion-free scan, median (ml) | 12.165 | 15.716 |
| Components per lesion-free scan, median | 2.000 | 3.000 |
| Predicted volume / pancreas volume, lesion-free, median | 15.8% | 20.8% |
| Scan-level AUC, maximum lesion probability | 0.468 | 0.446 |
| Scan-level AUC, predicted lesion volume | 0.587 | 0.545 |

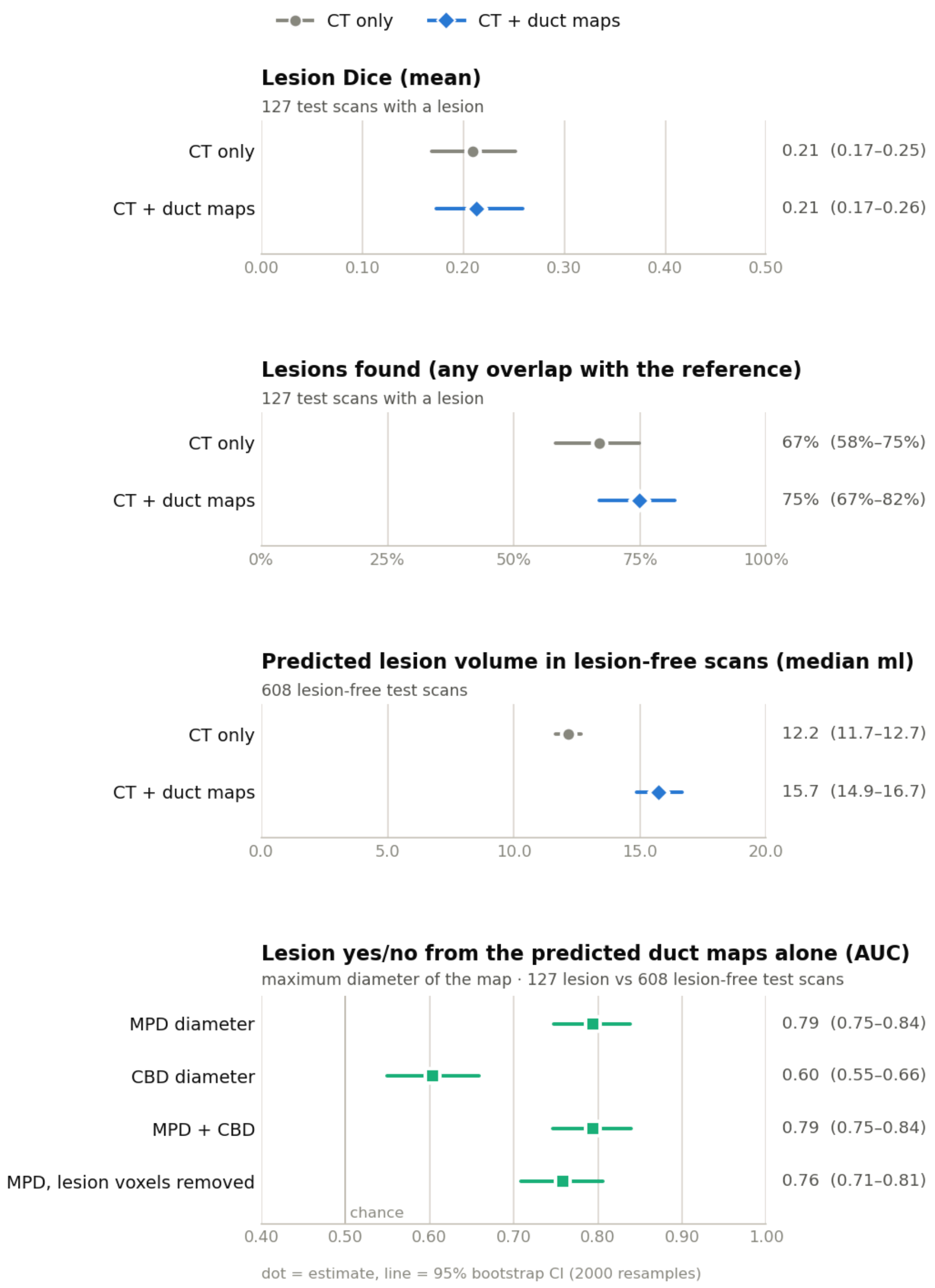

**Figure 2.** Test-set results (dot: estimate; line: 95% case-bootstrap CI). The upper three panels compare the two lesion models; the lowest panel shows scan-level lesion presence from the maximum diameter of the predicted maps alone (Section 5.5).

Figure 3 shows four test scans chosen by fixed rules.

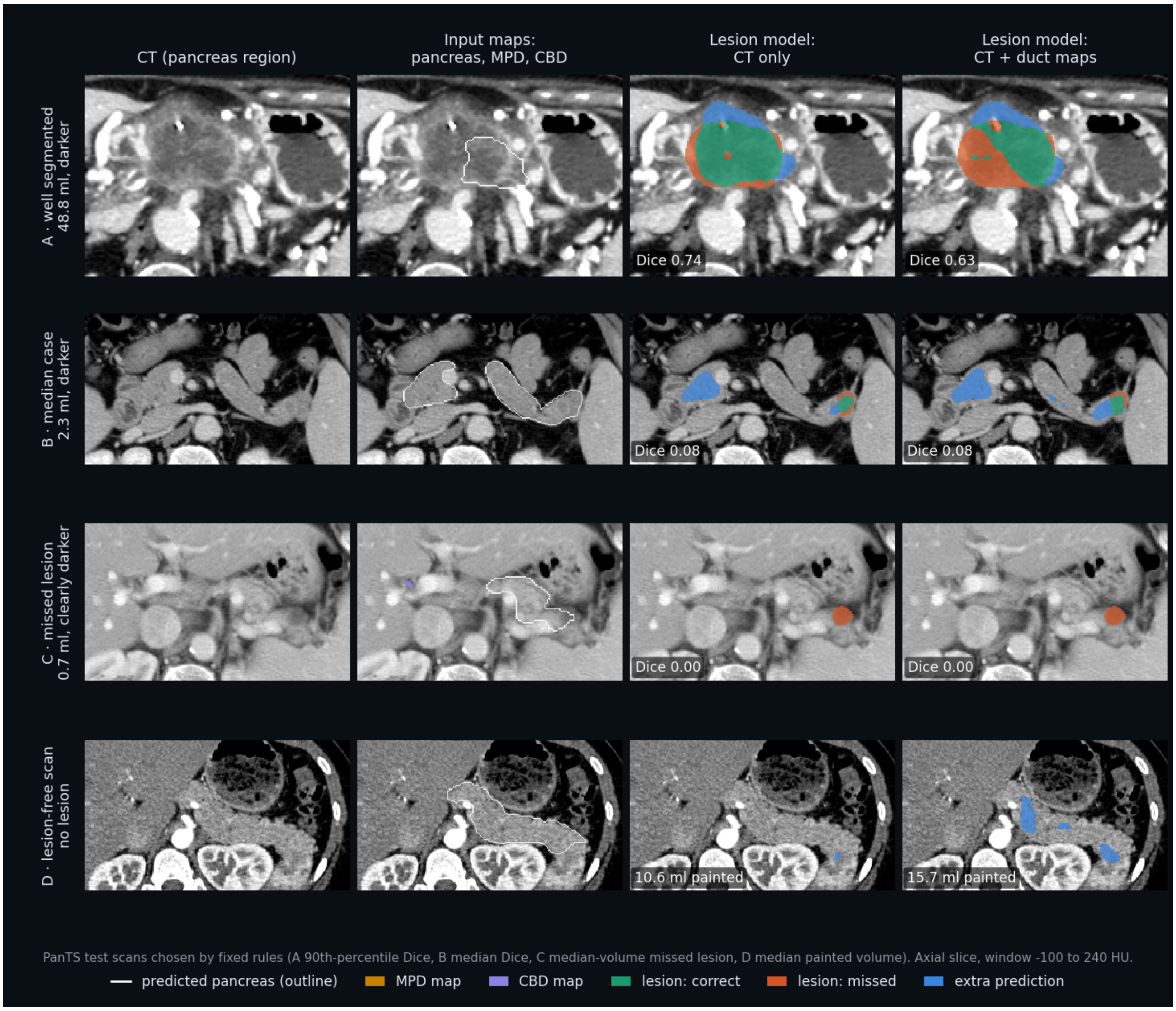

**Figure 3.** Four test scans chosen by fixed rules, not by eye: A nearest the 90th-percentile Dice of the model with duct maps, B nearest its median Dice, C the missed lesion with the median volume among missed lesions, D the lesion-free scan with the median predicted volume. Axial slice with the largest reference lesion (A-C) or prediction (D), window -100 to 240 HU. Columns: CT; input maps (predicted pancreas outline, MPD, CBD); lesion prediction of the CT-only model and of the model with duct maps (green: correct, orange: missed, blue: extra). Per-case values are from the test evaluation.

## 5.2 Where the Dice is lost (out of fold)

Out of fold, mean lesion Dice was 0.285; 150 of 168 lesions were touched by a prediction (89.3%), and 97.4% of lesion-free scans had a prediction. Removing every component that did not touch the lesion raised Dice only to 0.309. In contrast, if the touching components had contained only lesion, Dice would have reached 0.550. In the found cases the touching component measured a median 12.8 ml against a lesion volume of 5.9 ml; it covered a median 56.4% of the lesion, and 29.0% of its volume was lesion. The prediction was thus both too large and displaced. Its non-lesion part lay inside the predicted pancreas (median 96.2%), with 46.9% of it more than 5 mm from the lesion, and its intensity was close to that of normal pancreas (median difference -2.2 HU). Raising the probability threshold shrank the prediction but lowered Dice (0.272 at 0.5, 0.236 at 0.99, fold 0).

## 5.3 Attenuation and size

Table 2 stratifies the out-of-fold lesions by attenuation relative to normal pancreas. Lesions with similar attenuation (51 of 168) had the highest Dice (0.385); the found rate was high in both the similar and the darker strata (94.1% and 96.3%). Clearly darker lesions (44), whose voxels are separable from normal pancreas by intensity, had the lowest Dice (0.208) and coverage (median 16.6%). Inside the touching components, the lesion part was clearly darker than the rest in 15 of 142 measurable cases; of the 34 clearly darker lesions among them, 12 stayed clearly darker inside the component. Small lesions (≤ 2.15 ml) had Dice 0.075, against 0.320 and 0.460 for the middle and large tertiles. On the test set, clearly darker lesions again had the lowest Dice (0.088 (95% CI 0.047 to 0.133)) and coverage (median 2.1%), and they were more frequent (52 of 127); clearly brighter lesions were few (7) (Table 2, right). The clearly darker test lesions were smaller (median 2.9 ml, out of fold 5.6 ml) and darker (median 39 HU, out of fold 50 HU); 28.8% had a median attenuation below 30 HU (out of fold 20.5%). By size, test Dice with duct maps was 0.026, 0.191 and 0.428 for the small, middle and large out-of-fold tertiles.

**Table 2.** Lesion attenuation relative to normal pancreas. Out of fold: duct-map fold models. Test: duct-map model.

| Attenuation | Out of fold, n | Found | Dice | Coverage, median | Test, n | Test Dice (95% CI) |
|---|---|---|---|---|---|---|
| Clearly brighter (AUC ≤ 0.4) | 19 | 78.9% | 0.242 | 45.4% | 7 | 0.352 (95% CI 0.167 to 0.547) |
| Similar (0.4–0.6) | 51 | 94.1% | 0.385 | 68.3% | 25 | 0.333 (95% CI 0.242 to 0.430) |
| Darker (0.6–0.8) | 54 | 96.3% | 0.269 | 46.7% | 43 | 0.271 (95% CI 0.193 to 0.358) |
| Clearly darker (> 0.8) | 44 | 79.5% | 0.208 | 16.6% | 52 | 0.088 (95% CI 0.047 to 0.133) |

Figure 4 shows the test-set Dice of both lesion models by attenuation and by lesion size.

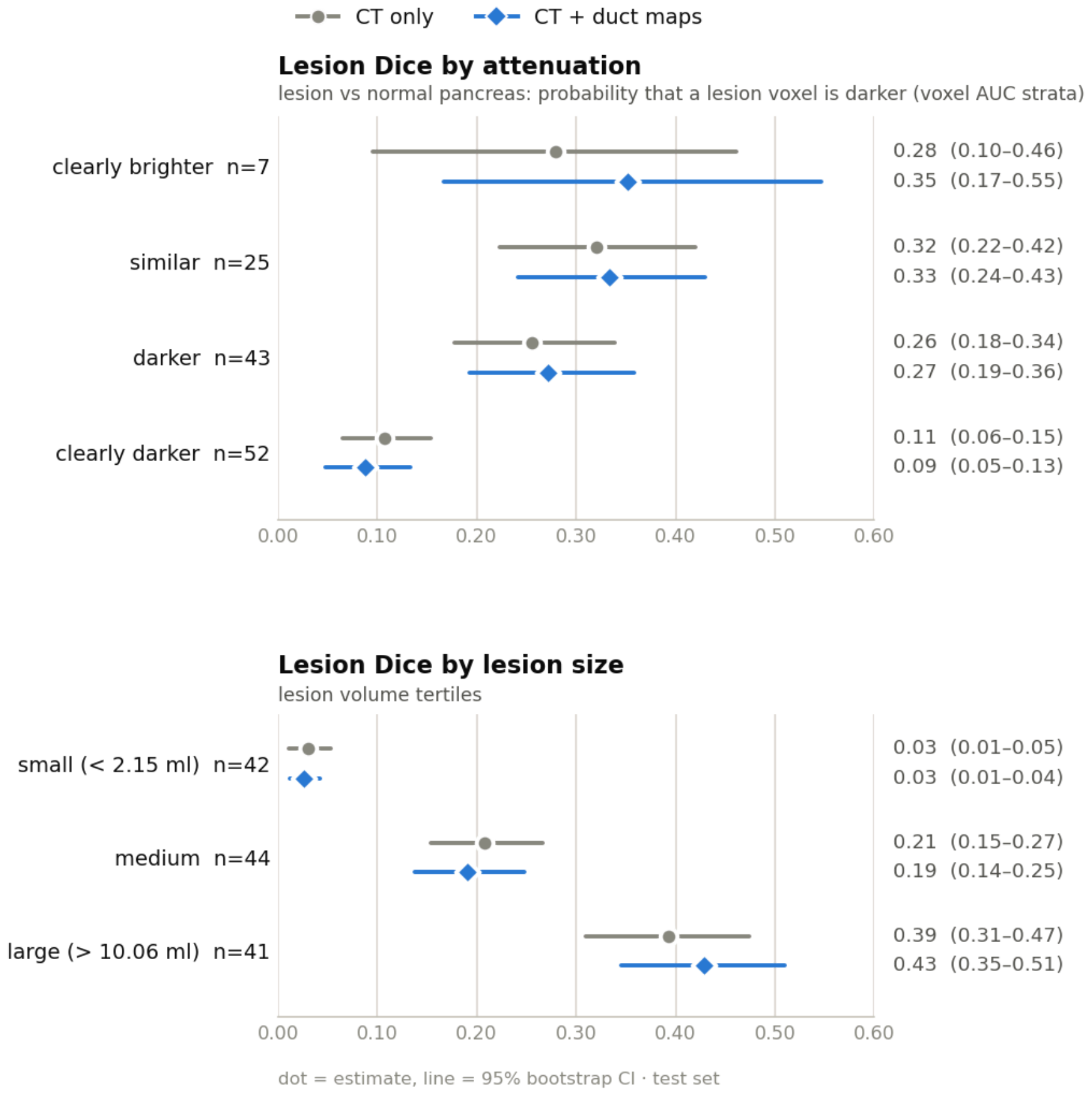

**Figure 4.** Test-set lesion Dice by attenuation stratum (voxel AUC of lesion versus normal pancreas) and by lesion-volume tertile, for both lesion models (dot: estimate; line: 95% case-bootstrap CI).

## 5.4 Candidate filtering

At operating points that kept 80% of lesion scans found, the share of lesion-free scans that still had a candidate was 87.4% for the crop CNN, 79.5% for the component volume and 74.5% for gradient boosting (Table 3). The scan-level AUC of the maximum candidate score was 0.631, 0.674 and 0.716. At the component level, precision at 80% recall was low for all three filters (Table 3). The crop CNN did not outperform the single volume feature.

**Table 3.** Candidate filtering, out of fold (168 lesion scans, 1433 lesion-free).

| Filter | Lesion-free scans with a candidate at 80% found | Scan-level AUC | Component precision at 80% recall (fold 1 / fold 0 candidates) |
|---|---|---|---|
| Component volume | 79.5% | 0.674 | 0.078 / 0.108 |
| Gradient boosting, 30 measurements | 74.5% | 0.716 | 0.070 / 0.074 |
| 3D ResNet-10 on CT crop | 87.4% | 0.631 | 0.067 / 0.065 |

### 5.5 Duct and CBD diameters as standalone signs

Table 4 and the lowest panel of Figure 2 show how well the predicted maps alone separated scans with a lesion from lesion-free scans. The MPD diameter reached a test AUC of 0.794 (95% CI 0.747 to 0.839) (out of fold 0.665 (95% CI 0.613 to 0.715)). The CBD diameter alone was weak (0.603 (95% CI 0.549 to 0.658)), and adding it to the MPD did not change the test AUC (difference -0.000 (95% CI -0.009 to +0.010)); the fitted weight of the CBD was small (standardised coefficients 0.759 for the MPD and 0.148 for the CBD). Part of the MPD signal came from map voxels on the lesion itself: in 24.4% of test lesion scans at least 10% of the predicted MPD lay inside the lesion, and removing these voxels lowered the test AUC to 0.758 (95% CI 0.708 to 0.806). The median MPD diameter was 4.8 mm in lesion scans and 3.2 mm in lesion-free test scans. By lesion size, the test AUC of the MPD diameter was 0.730, 0.878 and 0.767 for small, middle and large lesions (out of fold 0.601, 0.659 and 0.735). Out-of-fold maps came from a duct model trained on those scans, and the out-of-fold and test AUCs differed; we did not examine why.

**Table 4.** Scan-level lesion presence from the predicted maps alone (maximum diameter). MPD + CBD: logistic regression fitted on the out-of-fold scans; its out-of-fold AUC is in-sample.

| Sign | Out of fold AUC (95% CI) | Test AUC (95% CI) | Test AUC, lesion voxels removed | Test AUC by lesion size (small / middle / large) |
|---|---|---|---|---|
| MPD only | 0.665 (95% CI 0.613 to 0.715) | 0.794 (95% CI 0.747 to 0.839) | 0.758 | 0.730 / 0.878 / 0.767 |
| CBD only | 0.580 (95% CI 0.530 to 0.630) | 0.603 (95% CI 0.549 to 0.658) | 0.598 | 0.624 / 0.578 / 0.610 |
| MPD + CBD | 0.673 (95% CI 0.620 to 0.721) | 0.793 (95% CI 0.746 to 0.839) | 0.758 | 0.730 / 0.876 / 0.770 |

Table 5 gives fixed cutoffs. An MPD diameter of at least 5 mm was specific (test specificity 95.9%) but present in a minority of lesion scans (test sensitivity 44.1%). The predicted CBD was wide in many lesion-free scans (median 6.9 mm), so no CBD cutoff was both sensitive and specific.

**Table 5.** Sensitivity and specificity of fixed diameter cutoffs (out of fold 168 lesion and 1422 lesion-free scans; test 127 and 608).

| Cutoff | Out of fold sensitivity | Out of fold specificity | Test sensitivity (95% CI) | Test specificity (95% CI) |
|---|---|---|---|---|
| MPD ≥ 3 mm | 82.7% | 23.4% | 96.1% (95% CI 92.1%–99.2%) | 25.8% (95% CI 22.5%–29.4%) |
| MPD ≥ 4 mm | 57.1% | 68.1% | 66.1% (95% CI 58.3%–74.0%) | 73.4% (95% CI 70.1%–77.0%) |
| MPD ≥ 5 mm | 33.3% | 95.2% | 44.1% (95% CI 35.4%–52.8%) | 95.9% (95% CI 94.2%–97.4%) |
| MPD ≥ 6 mm | 22.6% | 98.0% | 33.9% (95% CI 26.0%–41.7%) | 97.4% (95% CI 96.1%–98.5%) |
| CBD ≥ 6 mm | 81.5% | 21.4% | 86.6% (95% CI 80.3%–92.1%) | 24.0% (95% CI 20.7%–27.6%) |
| CBD ≥ 8 mm | 44.6% | 67.2% | 41.7% (95% CI 33.1%–50.4%) | 72.9% (95% CI 69.2%–76.5%) |
| CBD ≥ 10 mm | 23.2% | 92.3% | 16.5% (95% CI 10.2%–23.6%) | 93.6% (95% CI 91.4%–95.6%) |
| CBD ≥ 12 mm | 10.1% | 97.7% | 6.3% (95% CI 2.4%–11.0%) | 98.2% (95% CI 97.0%–99.2%) |

## 6. Discussion

The breakdown points away from false components in lesion scans as the main cause of low Dice: even a perfect component filter would have added little, while the extent of the correctly located component carried most of the loss. The excess was ordinary-looking pancreas, often several millimetres away from the lesion, which explains why neither a higher threshold nor component-level classifiers helped.

The attenuation analysis separates two problems. About a third of lesions had the same attenuation as normal pancreas; for them the CT contains little boundary information and the low overlap is expected for any intensity-driven method. Clearly darker lesions, in contrast, are visible by intensity but were the least covered, out of fold and on the test set. The image therefore holds a boundary that the model did not use. We could not tell whether this is a learning problem or a matter of lesion type: PanTS does not separate cysts from solid lesions, and part of these lesions had near-water attenuation. It is a concrete target for future work.

Predicted duct maps as an extra input, as in [3], did not change lesion Dice on the test set. They made the lesion model predict more overall: more lesions were touched, and lesion-free scans received larger predictions, so the higher found rate did not improve scan-level separation.

Predictions in nearly all lesion-free scans are the main obstacle to using this pipeline for scan-level detection. We did not resolve whether this comes from the training setup (for example, how little lesion-free scans contribute to a batch Dice loss) or from the data; we leave it as an open problem.

**Limitations.** Lesion type is unknown in PanTS, so these results mix PDAC with other lesions and cysts. The split is by scan without patient identifiers. Out-of-fold analyses used two of five folds, and stages 1 and 2 were in-sample for those scans. The test set had previously been used to compare duct models. The duct and CBD diameter analyses were specified after the main test run; their only fitted part was trained on out-of-fold scans. Test Dice was lower than validation Dice (0.276 with duct maps, 16 lesions) and out-of-fold Dice; the test scans come from the PanTS ImageTe partition and the training scans from ImageTr. The partitions differ in case mix: venous-phase scans made up 77.3% of the test set against 50.0% of the training set (79.5% and 56.6% of lesion scans), and clearly darker lesions were more frequent; whether they also differ in source institution was not examined. There is no external test set and no reader study. The duct-map benefit was evaluated for one architecture and one training length.

## 7. Conclusion

On a large public CT dataset, an end-to-end pancreas → duct → lesion pipeline reached a test lesion Dice of 0.213; predicted duct maps as an extra input did not change it. The loss came mainly from the extent of correctly located predictions and from predictions in lesion-free scans. Iso-attenuating lesions bound what intensity-based boundaries can achieve, whereas clearly hypoattenuating lesions, the least covered both out of fold and on the test set, hold a boundary the model did not use and are the clearest target for improvement.

## Code and data availability

Code for preprocessing, training, inference and evaluation, together with the data splits, is available at https://github.com/Kang812/pants-lesion-failure-analysis. Re-running the released evaluation script on eight test scans reproduced the saved per-case values exactly. PanTS is available under its own licence (CC BY-NC-SA 4.0).